\documentclass[aps, prl, reprint, superscriptaddress]{revtex4-1}
\pdfoutput=1 
\usepackage{graphicx}
\usepackage{color,amsmath}

\usepackage[normalem]{ulem}

\usepackage[colorlinks,linkcolor=blue,citecolor=blue,urlcolor=blue]{hyperref}
\usepackage{siunitx}

\usepackage{comment}
\begin{document}

\title{Experimental Signatures of the Inverted Phase in InAs/GaSb Coupled Quantum Wells}

\author{Matija Karalic}
\email{makarali@phys.ethz.ch}
\affiliation{Solid State Physics Laboratory, ETH Zurich, 8093 Zurich, Switzerland}

\author{Susanne Mueller}
\affiliation{Solid State Physics Laboratory, ETH Zurich, 8093 Zurich, Switzerland}

\author{Christopher Mittag}
\affiliation{Solid State Physics Laboratory, ETH Zurich, 8093 Zurich, Switzerland}

\author{Kiryl Pakrouski}
\affiliation{Theoretische Physik and Station Q Zurich, ETH Zurich, 8093 Zurich, Switzerland}

\author{QuanSheng Wu}
\affiliation{Theoretische Physik and Station Q Zurich, ETH Zurich, 8093 Zurich, Switzerland}

\author{Alexey A. Soluyanov}
\affiliation{Theoretische Physik and Station Q Zurich, ETH Zurich, 8093 Zurich, Switzerland}
\affiliation{Department of Physics, St. Petersburg State University, St. Petersburg, 199034 Russia}

\author{Matthias Troyer}
\affiliation{Theoretische Physik and Station Q Zurich, ETH Zurich, 8093 Zurich, Switzerland}
\affiliation{Quantum Architectures and Computation Group, Microsoft Research, Redmond, WA, USA}
\affiliation{Microsoft Research Station Q, Santa Barbara, CA, USA}

\author{Thomas Tschirky}
\affiliation{Solid State Physics Laboratory, ETH Zurich, 8093 Zurich, Switzerland}

\author{Werner Wegscheider}
\affiliation{Solid State Physics Laboratory, ETH Zurich, 8093 Zurich, Switzerland}

\author{Klaus Ensslin}
\affiliation{Solid State Physics Laboratory,  ETH Zurich, 8093 Zurich, Switzerland}

\author{Thomas Ihn}
\affiliation{Solid State Physics Laboratory, ETH Zurich, 8093 Zurich, Switzerland}

\date{\today}

\begin{abstract}
Transport measurements are performed on InAs/GaSb double quantum wells at zero and finite magnetic fields applied parallel and perpendicular to the quantum wells. We investigate a sample in the inverted regime where electrons and holes coexist, and compare it with another sample in the non-inverted semiconducting regime. Activated behavior in conjunction with a strong suppression of the resistance peak at the charge neutrality point in a parallel magnetic field attest to the topological hybridization gap between electron and hole bands in the inverted sample. We observe an unconventional Landau level spectrum with energy gaps modulated by the magnetic field applied perpendicular to the quantum wells. This is caused by strong spin-orbit interaction provided jointly by the InAs and the GaSb quantum wells.
\end{abstract}


\maketitle

The peculiar band lineup of coupled InAs/GaSb quantum wells (QW) can lead to the coexistence of electrons and holes at the charge neutrality point \cite{lakrimi_minigaps_1997, cooper_resistance_1998}. According to a recent prediction \cite{liu_quantum_2008}, this material system displays a topological insulator phase, whose existence and strength can be tuned by gate voltages and which hosts the quantum spin Hall (QSH) state \cite{kane_z2_2005, kane_quantum_2005, bernevig_quantum_2006}.
The topological insulator phase arises if the band ordering is inverted and coupling between electron and hole states opens a hybridization gap at the charge neutrality point.
So far, several groups have performed experiments on InAs/GaSb QWs demonstrating edge transport \cite{suzuki_edge_2013, suzuki_gate-controlled_2015} and conductance quantization consistent with the QSH state \cite{knez_evidence_2011, knez_observation_2014, du_robust_2015, mueller_nonlocal_2015}. However, the aforementioned works have not shown spin polarization of edge currents, as has been done in HgTe QWs \cite{brune_spin_2012}. Indeed, recent experiments \cite{nichele_edge_2016, nguyen_decoupling_2016} have detected edge transport in the topologically trivial phase of InAs/GaSb QWs, providing room for an alternative explanation for edge transport in general in this system. Therefore, it is of utter experimental importance to clearly and consistently distinguish the two phases and to understand their properties under applied electric and magnetic fields.

Here we present experimental data from two samples of different InAs well widths, one of them in the topological regime (inverted band alignment and hybridization gap, topologically non-trivial phase, TI sample) and the other in the semiconducting regime (non-inverted band alignment and normal gap, topologically trivial phase, NI sample). Both samples have comparable mobilities. Our experiments unravel the energy dispersion qualitatively and the charge carrier occupations of electrons and holes quantitatively. Our data show unique fingerprints of the inverted bandstructure and band hybridization, such as a  suppression of the resistance peak at the charge neutrality point as a function of in-plane magnetic field, pronounced odd-integer filling factors in the quantum Hall effect as well as the detection of a highly non-linear Landau level spectrum caused by strong spin-orbit interactions. These features are absent in the non-inverted NI sample.


\begin{figure}
\includegraphics[width=\columnwidth]{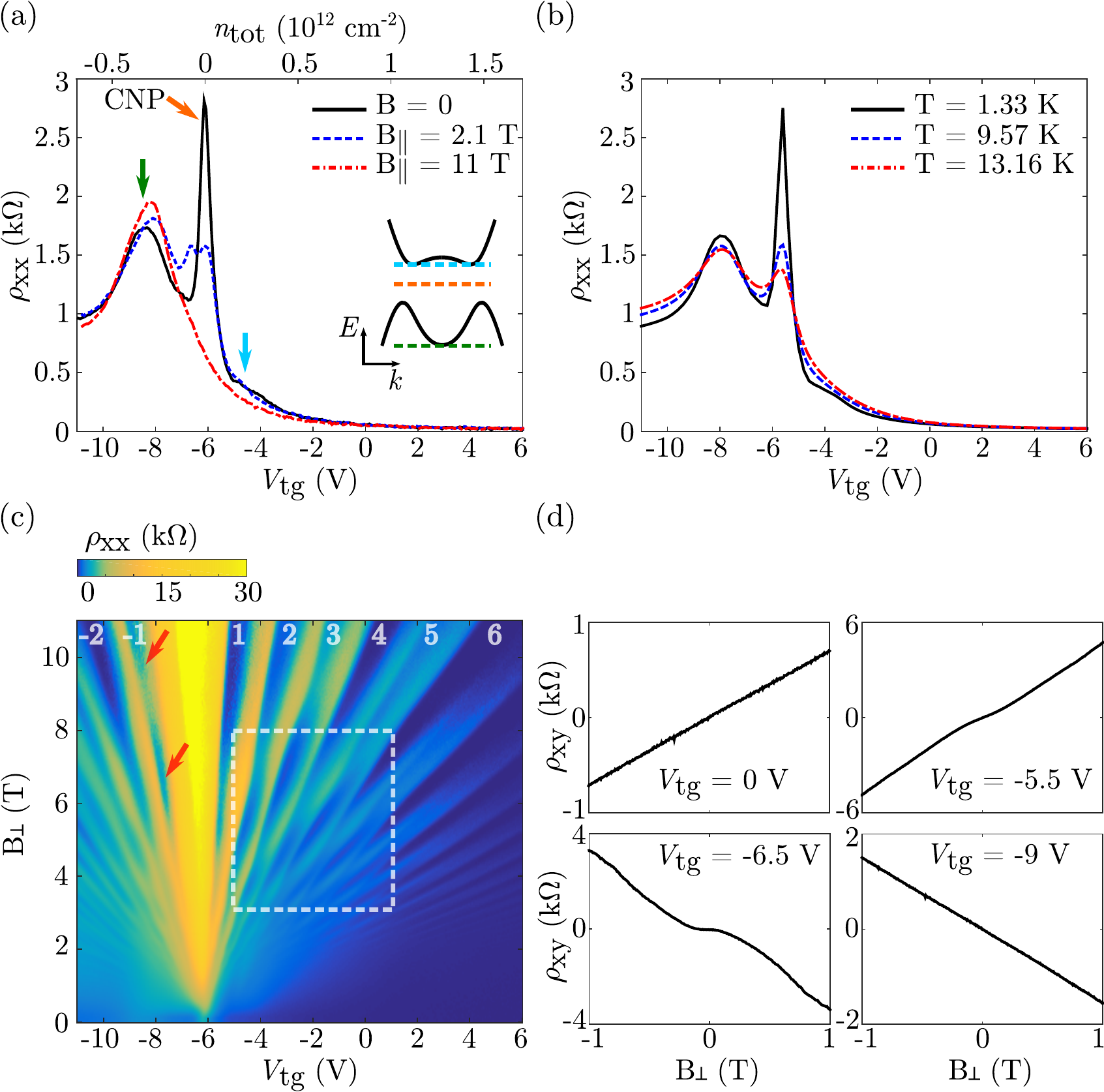}
\caption{\textbf{(a)} Longitudinal resistivity  $\rho_\mathrm{xx}$ of the TI sample as a function of top-gate voltage $V_\mathrm{tg}$, or equivalently, total density $n_\mathrm{tot}$, at different in-plane magnetic fields $B_{\parallel}$ at $T = 130$\,mK. The inset shows the position of the Fermi energy $E_\mathrm{F}$ at the points indicated by the arrows. \textbf{(b)} $\rho_\mathrm{xx}$ as a function of $V_\mathrm{tg}$ at different temperatures at zero magnetic field. The data was collected in a different cooldown compared to (a), resulting in a slight shift along the $V_\mathrm{tg}$-axis. \textbf{(c)} $\rho_\mathrm{xx}$ as a function of $V_\mathrm{tg}$ and perpendicular magnetic field $B_{\perp}$ at $T = 130$\,mK. The numbers indicate filling factors $\nu$, where we assign positive $\nu$ to electron and negative $\nu$ to hole Landau levels. The highlighted region is reproduced in Fig.\,\ref{fig2}. \textbf{(d)} Transverse resistivity $\rho_{xy}$ as a function of $B_{\perp}$ for 
$\left| B_{\perp} \right| \leq 1$\,T for different $V_\mathrm{tg}$ at $T = 130$\,mK.}  
\label{fig1}
\end{figure}

Details regarding wafer growth and sample processing are described in the  Supplemental Material \cite{supplement}.
Figure\,\ref{fig1}(a) shows the dependence of the longitudinal resistivity $\rho_\mathrm{xx}$ of the TI sample on top-gate voltage $V_\mathrm{tg}$. At zero in-plane magnetic field (solid line), we observe a sharp resistance peak at $V_\mathrm{tg}^\mathrm{CNP}=-6.08$\,V (red arrow) with a strong and broad satellite at $V_\mathrm{tg}=-8.4$\,V (green arrow) and a weak shoulder at $V_\mathrm{tg}=-4.5$\,V (blue arrow). From measurements of the total carrier density as a function of gate voltage [see Supplemental Material \cite{supplement} and the top axis in Fig.\,\ref{fig1}(a)] we identify the sharp central peak to coincide with the charge neutrality point (CNP). To the left (right) of this point the Fermi energy resides in the valence (conduction) band. In the past, structure in $\rho_\mathrm{xx}(V_\mathrm{tg})$ has only been observed in samples of high quality \cite{knez_finite_2010, qu_electric_2015, nichele_giant_2016} and been interpreted in terms of the influence of the energy-dependent density of states on $\rho_\mathrm{xx}$.

According to theory \cite{altarelli_electronic_1983, naveh_bandstructure_1995, liu_quantum_2008}, interband coupling mixes electron and hole states in the presence of band overlap and opens a gap in the energy spectrum. Figure\,\ref{fig1}(b) puts the existence of such a gap to the experimental test by showing the thermal activation of $\rho_\mathrm{xx}$. We observe that the resistance peak at charge neutrality decays strongly with the temperature increasing from $1.3$\,K to $13.2$\,K, indicating the possible presence of a hybridization gap. From the temperature where the peak has almost disappeared we qualitatively estimate an energy scale for the gap of a few meV. Comparing to Fig.\,\ref{fig1}(a) measured at $130$\,mK, we find that $\rho_\mathrm{xx}$ saturates below $1.3$\,K, compatible with disorder broadening of the gap at an energy scale of at least a few hundred $\SI{}{\micro\electronvolt}$.

We gain further experimental support for this interpretation by applying an in-plane magnetic field $B_\parallel$. The in-plane field induces a relative momentum shift $\Delta k = eB_{\parallel}d/\hbar$ between the electron and hole dispersions \cite{choi_anisotropic_1988, yang_evidence_1997}, where $d$ is the separation between electron and hole gases. In this way $B_{\parallel}$ turns the topological insulator into a semimetal once the variation of the gap center with direction becomes larger than the gap size. The corresponding measurement at $B_\parallel=2.1$\,T in Fig.\,\ref{fig1}(a) (dashed line) shows a strong suppression and a splitting of the resistance peak at the CNP as compared to $B_\parallel=0$. At $B_\parallel = 11$\,T (dash-dotted line) the peak has completely disappeared. This signature of the topological gap has not been seen in some earlier experiments in the inverted regime \cite{knez_finite_2010}, and a sharp resistance peak at the CNP together with a consistent in-plane magnetic field and temperature dependence have not been reported before in the relevant literature  \cite{knez_finite_2010, qu_electric_2015, nichele_giant_2016}. These measurements and their interpretation therefore represent the first of our two main experimental results.




At the CNP, the electron and hole densities, $n$ and $p$, are nonzero and equal. An interval in $V_\mathrm{tg}$ exists around this point, in which electrons and holes coexist due to the finite band overlap and both contribute to transport. We identify this region by analyzing the low-field transverse resistivity $\rho_\mathrm{xy}$ as a function of a magnetic field $B_\perp$ applied normal to the planes of the quantum wells. Representative examples of these measurements are shown in Fig.\,\ref{fig1}(d). While the curves at $V_\mathrm{tg}=0$ and $V_\mathrm{tg}=-9$\,V show linear behavior indicating pure electron and pure hole conduction, respectively, the two other traces exhibit nonlinearities around $B_\perp=0$ expected for the simultaneous contributions of electrons and holes. We analyze these nonlinearities quantitatively [see Supplemental Material \cite{supplement}] and find contributions of both carrier types to transport in the gate voltage range
$-8.7$ V $<V_\mathrm{tg}<$ $0$\,V. For $V_\mathrm{tg}<-9$\,V we identify the presence of two hole species of different densities and mobilities. This finding is a precursor of spin-splitting of the GaSb heavy hole band.

We complete the set of data presented for the TI sample with Fig.\,\ref{fig1}(c), depicting $\rho_\mathrm{xx}$ as a function of $V_\mathrm{tg}$ and $B_{\perp}$. This map shows the Landau fan of electrons for $V_\mathrm{tg}$ above the CNP, and that of holes below. Filling factors $\nu$ are indicated in the figure (negative for hole, positive for electron Landau levels).
A remarkable, regular pattern appears in the electron region above $B_{\perp}\approx 2$\,T. The minima in $\rho_\mathrm{xx}$ corresponding to integer filling factors $\nu$ are modulated in strength with magnetic field. A similar effect, although less pronounced, is visible on the hole side (at $\nu = -1$, see arrows). Other inverted samples we have measured show a stronger pattern on the hole side \cite{supplement}.

\begin{figure}
\includegraphics[width=\columnwidth]{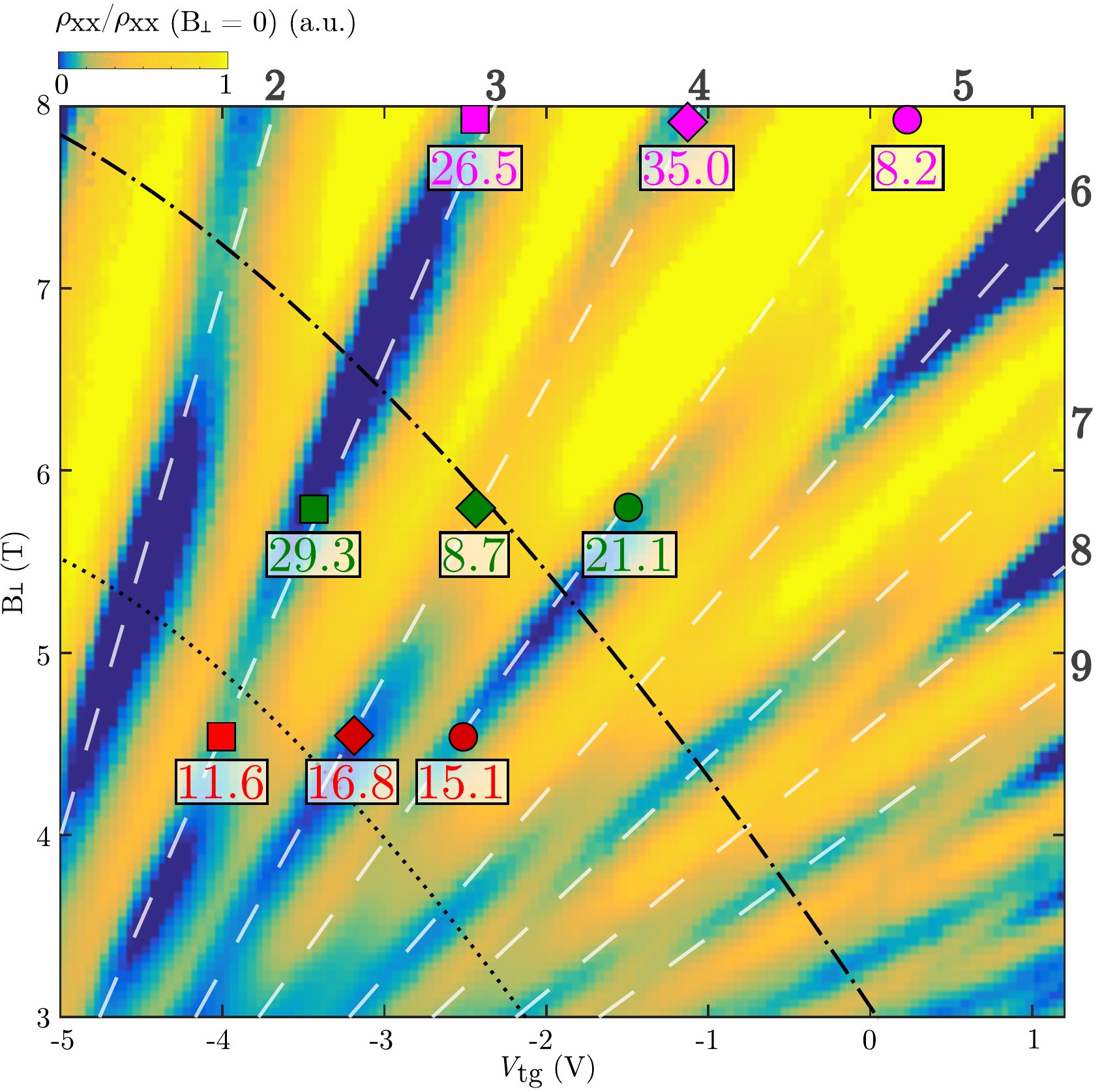}
\caption{Zoom-in of Fig.\,\ref{fig1}(c), where each row is normalized according to $\rho_\mathrm{xx}/\rho_\mathrm{xx} (B_{\perp} = 0)$. The numbers on the outside indicate filling factors $\nu$. The dashed lines are guides to the eye, marking the evolution of selected filling factors as a function of $V_\mathrm{tg}$ and $B_{\perp}$. Along the dotted (dash-dotted) line, even (odd) filling factors are enhanced and odd (even) filling factors are suppressed. The activation energy $\Delta$ at the positions of the symbols is found by fitting the temperature dependence of the minima of the longitudinal conductivity $\sigma_\mathrm{xx} = \rho_\mathrm{xx}/(\rho_\mathrm{xx}^{2} + \rho_\mathrm{xy}^{2})$ with an exponential function of the form $\sigma_{0}e^{-\Delta/(2k_{B}T)}$ \cite{note_fitting}. The unit of $\Delta$ is $\SI{}{\micro\electronvolt}$.}  
\label{fig2}
\end{figure}

A zoom into a region of the electron Landau fan is shown in Fig.\,\ref{fig2}. We study the modulation of the $\rho_\mathrm{xx}$ minima by measuring the associated activation energies for $\nu =3, 4, 5$ at the points marked in Fig.\,\ref{fig2}. They confirm that the largest activation energies lead to pronounced and deep minima in $\rho_\mathrm{xx}$, whereas the smallest lead to less pronounced and shallower minima. Therefore, the modulation of $\rho_\mathrm{xx}$ is caused by a modulation of energy gaps in the Landau level spectrum. This regular modulation is the second of the two main experimental results in this paper.

The values of the activation energies are unexpectedly small: with an electron mass of $m^{\star}_\mathrm{InAs} = (0.03\text{--}0.04)m_{0}$ \cite{kim_electron_1988}, the bare Landau level splitting is $\hbar e/m^\star_\mathrm{InAs} = 2.9\text{--}3.9$\,meV/T. The Zeeman splitting increases with magnetic field at a rate of $g^\star\mu_\mathrm{B} = 0.6\text{--}0.9$\,meV/T if we use $g^\star = 10\text{--}15$ \cite{brosig_inas-alsb_2000, mu_effective_2016}. The values for the activation energies in  Fig.\,\ref{fig2} are two orders of magnitude smaller than these expectations. We have also carried out tilted magnetic field measurements at fixed $B_{\perp}$ while varying $B_{\parallel}$ (data not shown). Unlike in a conventional two-dimensional electron system with Landau and Zeeman splitting only, we do not find a periodic suppression of even and odd filling factor minima in $\rho_\mathrm{xx}$ when $B_\parallel$ is increased (coincidence method \cite{fang_effects_1968}). Instead, $B_\parallel$ modulates the depth of $\rho_\mathrm{xx}$ minima, indicating that Landau gaps do not close, which we take as evidence for significant spin-orbit interaction mediated Landau-level coupling. In the recent Ref.\,\citenum{nichele_giant_2016}, the effects of strong spin-orbit interaction on the zero-field spin splitting were investigated, complementing our results obtained in the quantum Hall regime.



\begin{figure}
\includegraphics[width=\columnwidth]{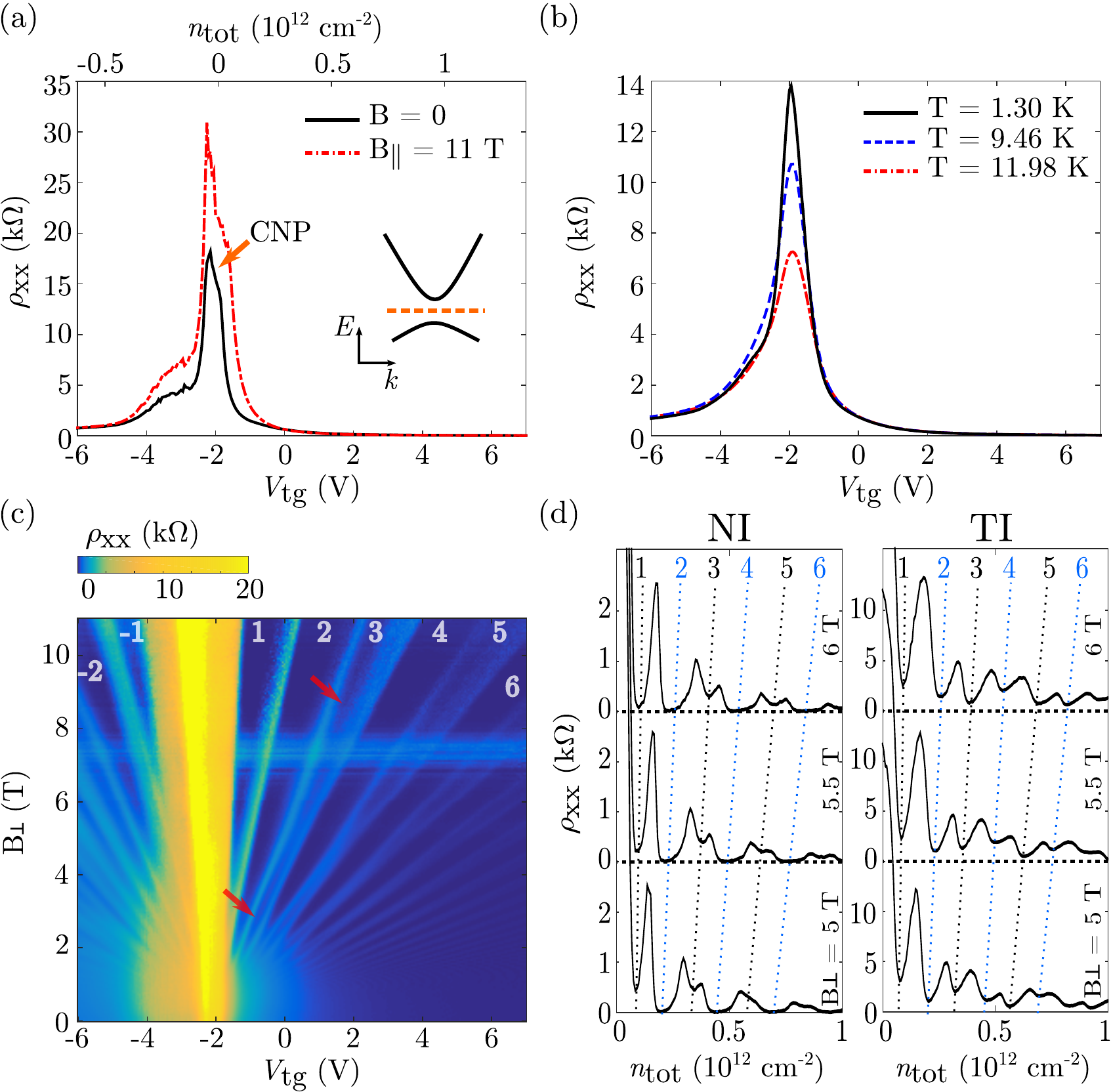}
\caption{\textbf{(a)} Longitudinal resistivity  $\rho_\mathrm{xx}$ of the NI sample as a function of top-gate voltage $V_\mathrm{tg}$, or equivalently, total density $n_\mathrm{tot}$, at zero in-plane magnetic field $B_{\parallel}$ and at $B_{\parallel} = 0$  at $T = 155$\,mK. The inset shows the position of the Fermi energy $E_\mathrm{F}$ in the trivial gap. \textbf{(b)} $\rho_\mathrm{xx}$ as a function of $V_\mathrm{tg}$ at different temperatures at zero magnetic field. The data was collected in a different cooldown compared to (a). \textbf{(c)} $\rho_\mathrm{xx}$ as a function of $V_\mathrm{tg}$ and perpendicular magnetic field $B_{\perp}$ at $T = 130$\,mK. The numbers indicate filling factors $\nu$, where we assign positive $\nu$ to electron and negative $\nu$ to hole Landau levels. The horizontal stripe visible around $B_{\perp} = 7$\,T is due to the sample being unstable. \textbf{(d)} Comparison of $\rho_\mathrm{xx}$ at $B_\perp = 5, 5.5, 6$\,T for NI and TI samples. The horizontal axis is the total density $n_\mathrm{tot}$. The dotted lines follow the movement of the denoted filling factors.}  
\label{fig4}
\end{figure}

We contrast the results presented above with measurements performed on the NI sample. Figs.\,\ref{fig4}(a--c) are the NI sample equivalent to the measurements shown in Fig.\,\ref{fig1}(a--c) for the TI sample. The top-gate voltage axis covers a density region comparable to that of the inverted sample.
The solid line in Fig.\,\ref{fig4}(a) depicts $\rho_\mathrm{xx}$ as a function of $V_\mathrm{tg}$ at zero magnetic field.  The charge neutrality point is identified with the sharp peak in the resistance around $V_\mathrm{tg}=-2$\,V (red arrow). The peak resistance is larger by more than a factor of six compared to the TI sample [Fig.\,\ref{fig1}(a)], indicating the presence of a larger gap in the NI sample.

Similar to the TI sample, the resistance peak decreases with increasing temperature, Fig.\,\ref{fig4}(b), in agreement with thermal activation across an energy gap between valence and conduction bands. However, the highest temperature of $12$\,K is not nearly sufficient to make the resistance peak vanish, in contrast to the behavior in the TI sample. The low magnetic field Hall effect is linear in $B_{\perp}$ even close to the CNP. This confirms the presence of only one charge carrier type near the band edges in the NI sample.

The relative momentum shift induced by $B_{\parallel}$ is not expected to cause any significant reduction of the gap in a non-inverted system. We could rather expect an increase of the gap and therefore an enhancement of the peak resistance at sufficiently high magnetic fields due to relative diamagnetic shifts of electron and hole states, which are related to their different spatial extent in growth direction and the different signs of curvature of the two dispersion relations. Comparing in Fig.\,\ref{fig4}(a) the dash-dotted line taken at $B_\parallel = 11$\,T with the solid zero field curve in the same figure, and with the corresponding pair of curves in Fig.\,\ref{fig1}(a), we indeed find this difference between topological and trivial insulator behavior.

We complete the comparison of the two samples in Fig.\,\ref{fig4}(c), showing $\rho_\mathrm{xx}$ as a function of $V_\mathrm{tg}$ and $B_{\perp}$ for the NI sample, in analogy to Fig.\,\ref{fig1}(c). The Landau fan branching out into the electron region resembles that of a conventional two dimensional electron system, where the energy spectrum in $B_{\perp}$ is determined by cyclotron and Zeeman energies. The fan lacks the strongly modulated energy gaps observed for the TI sample. Interestingly, even in the NI sample the minima of the odd filling factors oscillate weakly (as indicated exemplarily by the arrows), possibly due to the residual spin-orbit interaction intrinsic to the InAs QW. The cuts in Fig.\,\ref{fig4}(d) at constant $B_\perp$ highlight the difference between the NI and TI samples: while the NI sample shows a clear even-odd behavior with minima corresponding to even filling factors going to zero, the TI sample does not, exhibiting more complex structure.

We perform additional model-based analysis of the data to connect the experimental results with calculations of dispersion relations and Landau level spectra. The total density in our device follows the experimentally determined relation $n_\mathrm{tot}(V_\mathrm{tg}) = n-p = c(V_\mathrm{tg}-V_\mathrm{tg}^\mathrm{CNP})$ for the whole gate voltage range, with $c=1.41\times 10^{11}$\,cm$^{-2}$V$^{-1}$. Assuming parabolic dispersions for electrons and holes with effective masses $m^\star_\mathrm{InAs}=0.036m_{0}$ and $m^\star_\mathrm{GaSb}=0.36m_{0}$ \cite{kim_electron_1988}, we find an approximately linear decrease (increase) of the hole (electron) density with gate voltage. Under the same assumption the band overlap between the electron dispersion in InAs and the hole dispersion in GaSb also depends linearly on gate voltage. In Ref.\,\citenum{qu_electric_2015} the validity of such a model incorporating screening effects via quantum capacitances was shown.

A Fourier analysis of the Landau fan in Fig.\,\ref{fig1}(c) (discussed in the Supplemental Material \cite{supplement}) combined with the low-field Hall effect measurements lead to an estimate of a depletion voltage of holes (electrons) $V_\mathrm{tg}^{(\mathrm{p})}=2$\,V ($V_\mathrm{tg}^{(\mathrm{n})}=-9$\,V). Using these values we find the electron density at charge neutrality to be about $3\times 10^{11}$\,cm$^{-2}$. The band overlap increases linearly with gate voltage from approximately $3$\,meV at $V_\mathrm{tg}=-9$\,V to 76\,meV at $V_\mathrm{tg}=2$\,V. These numbers imply that the band alignment becomes non-inverted for $V_\mathrm{tg}<-10$\,V. 

Apart from the CNP peak in Fig.\,\ref{fig1}(a), we also see a prominent peak in the hole region (green arrow). This peak, which is only weakly affected by in-plane field and temperature, coincides roughly with the low-voltage end of the band overlap region. It occurs when $E_\mathrm{F}$ is close to the bottom of the conduction band (refer to inset of Fig.\,\ref{fig1}(a)). Minima of $\rho_\mathrm{xx}$ like that between this peak and the CNP peak, or the little kink of $\rho_\mathrm{xx}$ on the electron side (between the red and blue arrows) have been attributed in  Ref.\,\citenum{knez_finite_2010} to van Hove singularities in the density of states at band extrema occurring at nonzero wave vectors. The shoulder in the electron region is seen more clearly in other samples \cite{supplement}.

\begin{figure}
\includegraphics[width=\columnwidth]{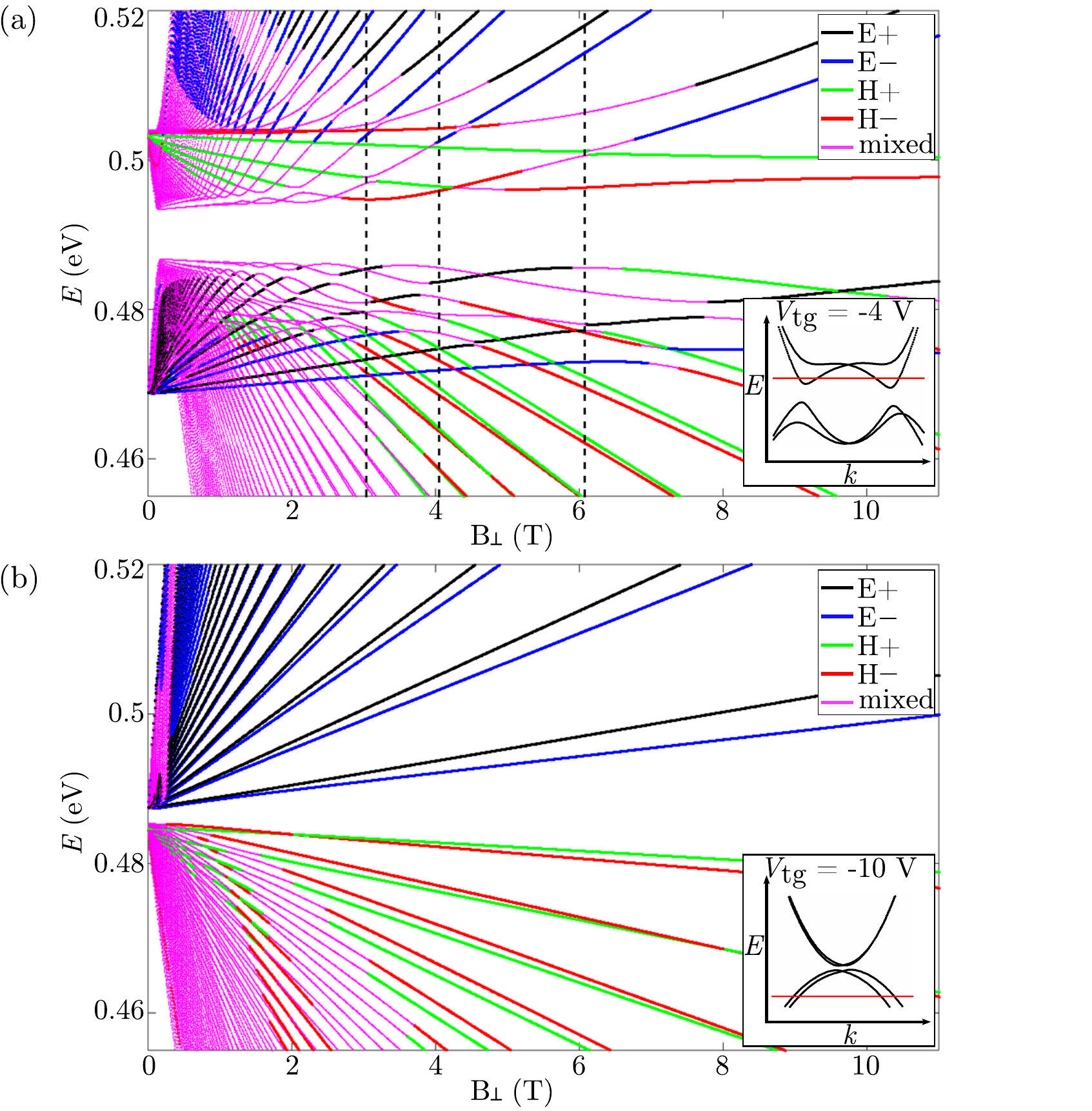}
\caption{\textbf{(a)} Landau level spectrum calculated numerically for the TI sample at $V_\mathrm{tg} = -4$\,V, inverted regime. States dominated by electrons or holes of certain spin and mixed states are shown in corresponding colors. The vertical dashed lines indicate the magnetic fields for $\nu = 2,3 ,4$ where the Fermi energy jumps to the next unoccupied Landau level. The inset depicts the dispersion Y--$\Gamma$--X and the position of the Fermi energy at $B_{\perp} = 0$. \textbf{(b)} As in (a), but at $V_\mathrm{tg} = -10$\,V, non-inverted regime. The spectrum calculated for the NI sample is qualitatively the same.}  
\label{figtheory}
\end{figure}

With the above insights in mind we discuss the origin of the remarkable pattern in the Landau fan seen in Figs.\,\ref{fig1}(c) and \ref{fig2} from a theoretical perspective. Fig.\,\ref{figtheory} shows a comparison of numerically calculated Landau level spectra for the inverted (a) and non-inverted regime (b) of the TI sample. These calculations are based on a $\vec{k}\cdot\vec{p}$ Hamiltonian with up to quadratic terms in $\vec{k}$ in the subspace of conduction band states in InAs and heavy hole valence band states in GaSb [see Supplemental Material \cite{supplement}]. Fig.\,\ref{figtheory}(b), calculated for $V_\mathrm{tg}=-10$\,V (non-inverted regime) shows the two Landau fans for electrons and holes modified by Zeeman splitting and spin-orbit interaction effects. Electron and hole states do not mix appreciably. The resulting Landau fans correspond to the observation on the NI sample shown in Fig.\,\ref{fig4}(c).

This behavior is in contrast to the Landau level spectrum at $V_\mathrm{tg}=-4$\,V (inverted regime) shown in Fig.\,\ref{figtheory}(a). Due to the inversion, the electron Landau fan starts off at lower energy than that of holes, and the two fans inevitably intersect. The off-diagonal terms in the Hamiltonian describing ordinary inter-band and spin-orbit inter-band coupling effects lead to avoided crossings of levels. These are the same terms that govern the appearance of the topological insulator gap at zero magnetic field. Due to the presence of hole-like Landau levels in the electron regime above the gap the spectrum is significantly denser than expected for pure InAs electron systems, in agreement with the smaller gaps observed in the experiment. Furthermore, if transitions of the Fermi energy between Landau levels at integer filling factors happen close to avoided crossing points of Landau levels, the gaps can be particularly small. These insights give a complete picture that is in qualitative agreement with the measurement presented in Fig.\,\ref{fig1}(c). Comparison between experiment and theory at a more quantitative level gives satisfactory results, as described in the Supplemental Material \cite{supplement}. Note that while in a transport measurement the gate voltage dependence of the Landau levels is linear independent of their energy spectrum, here the Landau level gap for a given filling factor is non-monotonic as a function of magnetic field, in agreement with Fig.\,\ref{figtheory}. This and the tilted-field measurements are why we conclude that the Landau level energies are non-linear as a function of magnetic field.


\begin{acknowledgments}
The authors acknowledge the support of the ETH FIRST laboratory and the financial support of the Swiss Science Foundation (Schweizerischer Nationalfonds, NCCR MARVEL and QSIT). KP, AS, QW and MT were supported by Microsoft Research.
\end{acknowledgments}


\begin{thebibliography}{18}
\bibitem{note_fitting} $\sigma_\mathrm{xx}$ in the smallest gaps, $\nu=4$ at $B_\perp=5.7$\,T and $\nu=5$ at $B_\perp=8$\,T, can be fit equally well with $\sigma_{0}e^{-(\Delta/(2k_{B}T))^{1/2}}$, but 
 the resulting $\Delta$ is within 10\% of those values given in Fig.\,\ref{fig2}. 
\end{thebibliography}

%

\end{document}